\documentclass{aa} 

\usepackage{natbib}
\usepackage{txfonts}
\usepackage{amsmath,amstext}
\usepackage{graphicx}
\usepackage{xcolor}
\usepackage{epstopdf}
\usepackage{float}
\usepackage{array}
\usepackage{mathtools}
\usepackage{booktabs}
\usepackage{subfigure}
\usepackage{url}
\usepackage{helvet}
\usepackage{tabularx}
\usepackage{multirow}
\usepackage[flushleft]{threeparttable}
\usepackage{lscape}
\usepackage{pdflscape}
\usepackage{longtable}
\usepackage{wasysym}
\usepackage{float}
\usepackage{relsize}
\usepackage{color}
\usepackage[breaklinks=true]{hyperref} 
\usepackage{breqn}
\usepackage{bm}

\hypersetup{colorlinks,breaklinks, linkcolor=blue,urlcolor=blue, anchorcolor=blue,citecolor=blue}

\def\kms{km s$^{-1}$}         
\def\gcm3{\hbox{g cm$^{-3}$}}       
\def\Msun{\hbox{$\mathrm{M}_{\astrosun}$}}             
\def\Rsun{\hbox{$\mathrm{R}_{\astrosun}$}}

\def\Mearth{\hbox{$\mathrm{M}_{\oplus}$}}
\def\Rearth{\hbox{$\mathrm{R}_{\oplus}$}}
\def\degr{\hbox{$^\circ$}}
\def\teff{T$_{\rm eff}$}
\def\logg{log~{\it g}}
\def\met{[Fe/H]}

\bibpunct{(}{)}{;}{a}{}{,}

\begin{document} 

   
      \title{Water content trends in K2-138 and other low-mass multiplanetary systems\thanks{Based on observations made with ESO Telescopes at the La Silla Paranal Observatory under programme ID 198.C-0.168.}}

   \author{L. Acuña\inst{1}
          \and T. A. Lopez\inst{1}
          \and T. Morel\inst{2}
          \and M. Deleuil \inst{1}
          \and O. Mousis \inst{1}
          \and A. Aguichine \inst{1}
          \and E. Marcq \inst{3}
          \and A. Santerne \inst{1}
          }

   \institute{Aix Marseille Univ, CNRS, CNES, LAM, Marseille, France \\ \email{lorena.acuna@lam.fr}
        \and
             Space sciences, Technologies and Astrophysics Research (STAR) Institute, Universit\'e de Li\`ege, Quartier Agora, All\'ee du 6 Ao\^ut 19c, B\^at. B5C, B4000-Li\`ege, Belgium
         \and
             LATMOS/CNRS/Sorbonne Université/UVSQ, 11 boulevard d'Alembert, Guyancourt, F-78280, France
        }

   \date{Received 5 October 2021; accepted 26 January 2022}
   
\abstract
   {Both rocky super-Earths and volatile-rich sub-Neptunes have been found simultaneously in multiplanetary systems, suggesting that these systems are appropriate to study different composition and formation pathways within the same environment.}   
   {We perform a homogeneous interior structure analysis of five multiplanetary systems to explore the compositional trends and its relation with planet formation. For one of these systems, K2-138, we present revised masses and stellar host chemical abundances to improve the constraints on the interior composition of its planets.}
   {We conduct a line-by-line differential spectroscopic analysis on the stellar spectra of K2-138 to obtain its chemical abundances and the planetary parameters. We select multiplanetary systems with five or more low-mass planets ($M$ < 20 $M_{\oplus}$) that have both mass and radius data available. We carry out a homogeneous interior structure analysis on the planetary systems K2-138, TOI-178, Kepler-11, Kepler-102 and Kepler-80. We estimate the volatile mass fraction of the planets in these systems assuming a volatile layer constituted of water in steam and supercritical phases. Our interior-atmosphere model takes into account the effects of irradiation on the surface conditions.}
   {K2-138 inner planets present an increasing volatile mass fraction with distance from its host star, while the outer planets present an approximately constant water content. This is similar to the trend observed in TRAPPIST-1 in a previous analysis with the same interior-atmosphere model. The Kepler-102 system could potentially present this trend. In all multiplanetary systems, the low volatile mass fraction of the inner planets could be due to atmospheric escape while the higher volatile mass fraction of the outer planets can be the result of accretion of ice-rich material in the vicinity of the ice line with later inward migration. Kepler-102 and Kepler-80 present inner planets with high core mass fractions which could be due to mantle evaporation, impacts or formation in the vicinity of rocklines.
}   
   {}


\maketitle
%

\section{Introduction}\label{intro}

Multiplanetary systems appear to be suitable distant laboratories to explore the diversity of small planets, and their formation and evolution pathways. This is the case of Kepler-36 \citep{2012Sci...337..556C}, where its two planets, b and c, present periods of 14 and 16 days with densities of 7.5 and 0.9 \gcm3, respectively. This suggests that these planets may have formed in different environments within the same protoplanetary disk before migrating inwards. Furthermore, a decreasing density gradient with distance from the host star in multiplanetary systems with 6 to 7 planets, such as TRAPPIST-1 \citep{2021arXiv210108172A,agol21} and TOI-178 \citep{Leleu21} suggest that there might be a transition between the rocky, inner super-Earths and the outer, volatile-rich sub-Neptunes. This transition is most probably due to the presence of the snowline in the protoplanetary disk \citep{Ruden99}.

Nevertheless, there are presently several limitations to determine the variation of the volatile mass fraction of planets within their systems that include the precision reached on the fundamental parameters of both the planets and the star, and the different assumptions considered between different interior structure models. These assumptions include whether the volatile layer of the planet is fully constituted of H/He \citep{lopez_fortney14}, an ice layer \citep{Zeng19}, an ice layer with a H/He atmosphere on top \citep{dorn15} or a steam and/or supercritical water layer \citep{mousis20,Turbet20}. To overcome the differences in volatile mass fraction estimates of multiplanetary systems due to the different compositions of the volatile layer between interior structure models, we perform a homogeneous analysis of the interior structure and composition of several multiplanetary systems. In our interior structure model, we consider that the volatile layer is water-dominated, following the approach of \cite{mousis20} and \cite{2021arXiv210108172A}. This analysis allows us to uncover volatile and core mass fraction trends, and their connection with planet formation and evolution. 
We use already published masses, radii and stellar composition data for four systems, and perform our own spectroscopic analysis to improve the parameters of one system, K2-138, whose detection was reported in \citet{christiansen18}. K2-138 harbours six small planets in a chain of near 3:2 mean-motion resonances and benefited of a radial velocity ground-based follow-up with HARPS on the 3.6m telescope at La Silla Observatory, leading to the confirmation and mass measurements of the four inner planets \citep{2019A&A...631A..90L}, with relatively good precisions given the standard today. In order to bring stronger constraints on the stellar parameters and abundances, and further reduce the degeneracies in the planetary structure modelling, we carried out an in-depth analysis of K2-138.

Section \ref{sect_spectroscopic_analysis} presents the new detailed analysis of the stellar host in the K2-138 system, which allowed us to derive stellar fundamental parameters and the elemental abundances, using the Sun and $\alpha$ Cen B as benchmarks. Section \ref{sect:pastis} describes a new Bayesian analysis of the HARPS radial velocities and K2 photometry, using the new stellar parameters. 

We describe our interior-atmosphere modelling in Sect. \ref{sect:methodology}, including our calculation of atmospheric mass-loss rates to infer the current presence or absence of volatiles. We present the volatile and core mass fraction trends for each mutiplanetary system as a result of our homogeneous analysis in Sect. \ref{sect:results}. Finally, we discuss the planet formation and evolution mechanisms that could have shaped these compositional trends in Sect. \ref{sect:discussion}. We present our concluding remarks in Sect. \ref{sect:conclusion}.

\section{Spectroscopic analysis}\label{sect_spectroscopic_analysis}
\object{K2-138} stellar parameters and abundances were derived based on a differential, line-by-line analysis relative to the Sun. The solar abundances are determined as part of such an analysis \citep[e.g.][]{Melendez12} and a set of reference values is not assumed. We used the HARPS spectra retrieved under programme ID 198.C-0.168. These were corrected from systemic velocity and planetary reflex-motion, removing the spectra with a S/N lower than ten in order 47 (550 nm) and the ones contaminated by the moonlight (S/N above 1.0 in fibre B). We then co-added the spectra in a single 1D spectrum and normalised it to the continuum. For the Sun, we used the HARPS spectra extracted from the ESO instrument archives\footnote{\url{http://archive.eso.org}}, acquired under programme ID 088.C-0323. The reduction of the solar spectrum, obtained as the spectrum of the light reflected by Vesta, is detailed in \citet{2016MNRAS.457.3637H} and the co-addition was performed as for \object{K2-138}. 

The stellar parameters and abundances of 24 metal species were self-consistently determined from the spectra, plane-parallel MARCS model atmospheres \citep{gustafsson08}, and the 2017 version of the line-analysis software MOOG originally developed by \citet{sneden73}.

The equivalent widths (EWs) were measured manually using IRAF\footnote{{\tt IRAF} is distributed by the National Optical Astronomy Observatories, operated by the Association of Universities for Research in Astronomy, Inc., under cooperative agreement with the National Science Foundation.} tasks assuming Gaussian profiles. Strong lines with RW = $\log$ (EW/$\lambda$) $>$ --4.80 were discarded. This constraint on the line strength was relaxed for Mg because it would result in no \ion{Mg}{i} lines left.

\subsection{Stellar parameters}\label{sect_stellar_parameters}
The stellar parameters of \object{K2-138} and \object{$\alpha$ Cen B} appear to be similar (see below). Therefore, we also analysed the latter for benchmarking because it has accurate and nearly model-independent $T_{\rm eff}$ and $\log g$ estimates from long-baseline interferometry and asteroseismology, respectively. \object{K2-138} and \object{$\alpha$ Cen B} were observed with exactly the same instrumental set-up, which ensures the highest consistency \citep{bedell14}. The \object{$\alpha$ Cen B} spectra were selected from the ESO archive, keeping those corrected from the Blaze and with S/N higher than 350 in order 47. For \object{$\alpha$ Cen B}, we adopt in the following $T_{\rm eff}$ = 5231$\pm$21 K derived by \citet{kervella17a} from their VLTI/PIONIER measurements and the bolometric flux of \citet{boyajian13}. We also assumed $\log g$ = 4.53$\pm$0.02 dex \citep{heiter15} based on scaling relations making use of the frequency of maximum oscillation power, $\nu_{\rm max}$, determined from radial-velocity time series by \citet{kjeldsen08}.

The model parameters ($T_{\rm eff}$, $\log g$, $\xi$, and [Fe/H]) were iteratively modified until the excitation and ionisation balance of iron is fulfilled and the \ion{Fe}{i} abundances exhibit no trend with RW. The abundances of iron and the $\alpha$ elements were also required to be consistent with the values adopted for the model atmosphere. For the solar analysis, $T_{\rm eff}$ and $\log g$ were held fixed to 5777 K and 4.44 dex, respectively, whereas the microturbulence, $\xi$, was left as a free parameter. The uncertainties in the stellar parameters were computed as in \citet{morel18}. 

We first carried out the analysis of \object{$\alpha$ Cen B} and \object{K2-138} using various iron line lists \citep{biazzo12,doyle17,feltzing01,jofre14,melendez14,morel14,reddy03,tsantaki19}. For \citet{jofre14}, we adopted their FGDa line list. The goal is to identify the line list that provides the most accurate parameters based on a comparison with the interferometric and asteroseismic constraints at hand for \object{$\alpha$ Cen B}. To ensure the highest consistency, the spectral features the analysis is based on for a given line list were exactly the same for the three stars.

\begin{figure*}[h!]
\centering
\includegraphics[trim=0 180 0 90,clip,width=0.8\hsize]{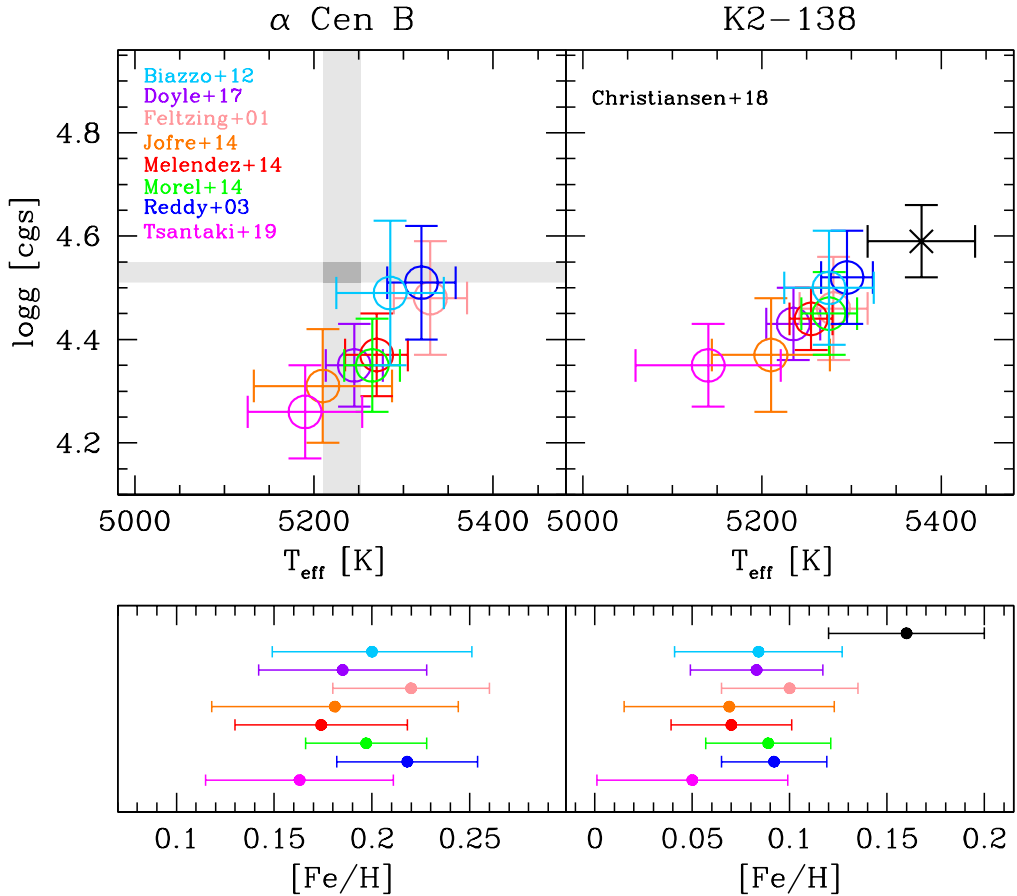}
\caption{Results of the analysis of \object{$\alpha$ Cen B} (left panels) and \object{K2-138} (right panels) using the various iron line lists. The colour coding for each line list is indicated in the upper left panel. The parameters of \object{K2-138} determined by \citet{christiansen18} are shown in the right panels. The grey shaded areas for \object{$\alpha$ Cen B} delimit the interferometric $T_{\rm eff}$ and seismic $\log g$ values ($\pm$1 $\sigma$; see Sect.~\ref{sect_stellar_parameters} for details).}
\label{fig_spectroscopic_parameters}
\end{figure*}

The parameters obtained are given in Table~\ref{tab_spectroscopic_parameters} and shown in Fig.~\ref{fig_spectroscopic_parameters}. The surface gravity of \object{$\alpha$ Cen B} appears to be underestimated in most cases. We also experimented with the LW13 Ti line list of \citet{tsantaki19} to constrain this quantity through Ti ionisation balance. As discussed by these authors, this leads to a larger value amounting here to $\sim$0.11 dex. However, it still falls short of matching the seismic value. As can be seen in Fig.~\ref{fig_spectroscopic_parameters}, the only notable difference between the parameters of \object{$\alpha$ Cen B} and \object{K2-138} is that the latter is slightly poorer in metals. Indeed, a differential analysis of \object{K2-138} with respect to \object{$\alpha$ Cen B} adopting the line list of \citet{biazzo12} gives the following results: $\Delta$$T_{\rm eff}$ = --10$\pm$45 K, $\Delta$$\log g$ = +0.02$\pm$0.09 dex, $\Delta$$\xi$ = +0.03$\pm$0.09 km s$^{-1}$ and $\Delta$[Fe/H] = --0.11$\pm$0.04. For the abundance analysis of \object{K2-138}, we adopt in the following the parameters provided by the line list of \citet{biazzo12}: $T_{\rm eff}$ = 5275$\pm$50 K, $\log g$ = 4.50$\pm$0.11, $\xi$ = 0.95$\pm$0.10 km s$^{-1}$ and [Fe/H] = +0.08$\pm$0.05. This choice is motivated by the fact that it leads to parameters that reproduce the reference ones of \object{$\alpha$ Cen B} within the errors. In addition, the metallicity is within the range of accepted values for the binary system \cite[][and references therein]{morel18}.

However, from the comparison to the interferometric-based $T_{\rm eff}$ in Fig.~\ref{fig_spectroscopic_parameters}, we cannot rule out that the effective temperature of \object{K2-138} is slightly overestimated at the $\sim$50 K level. The analysis was also repeated using Kurucz atmosphere models \citep{castelli03} The following modest deviations with respect to the default values (Kurucz -- MARCS) were found: $\Delta T_{\rm eff}$ $\sim$ +10 K, $\Delta \log g$ $\sim$ +0.02 dex, and $\Delta$[Fe/H] $\sim$ +0.02 dex. We will examine the robustness of our abundance results against such putative systematic errors in Sect.~\ref{sect_stellar_abundances}. In any case, we find that \object{K2-138} is cooler and less metal rich than concluded by \citet{christiansen18}. 

\begin{table*}[h!]
\caption{Stellar parameters of \object{$\alpha$ Cen B} and \object{K2-138}, as obtained from the various iron line lists. For iron, 42 Fe I and 4 Fe II lines were used.}
\label{tab_spectroscopic_parameters} 
\small
\centering
\begin{tabular}{l|cccc|cccc}
\hline\hline
                   & \multicolumn{4}{c}{\object{$\alpha$ Cen B}} & \multicolumn{4}{c}{\object{K2-138}} \\
                   & $T_{\rm eff}$ &  $\log g$     & $\xi$           &  [Fe/H]        &     $T_{\rm eff}$ &  $\log g$    &    $\xi$       &     [Fe/H]       \\
Iron line list     & [K]             &               & [km s$^{-1}$]   &                 &    [K]             &               &  [km s$^{-1}$]  &                 \\
\hline                                                                                                                                            
\citet{biazzo12}   & 5285$\pm$60     & 4.49$\pm$0.14 & 0.909$\pm$0.121 & 0.200$\pm$0.051 &   5275$\pm$50      & 4.50$\pm$0.11 & 0.945$\pm$0.099 & 0.084$\pm$0.043\\
\citet{doyle17}    & 5245$\pm$32     & 4.35$\pm$0.08 & 0.490$\pm$0.146 & 0.185$\pm$0.043 &   5235$\pm$30      & 4.43$\pm$0.07 & 0.450$\pm$0.146 & 0.083$\pm$0.034\\
\citet{feltzing01} & 5330$\pm$41     & 4.48$\pm$0.11 & 0.890$\pm$0.100 & 0.220$\pm$0.040 &   5280$\pm$38      & 4.46$\pm$0.10 & 0.915$\pm$0.084 & 0.100$\pm$0.035\\
\citet{jofre14}    & 5210$\pm$77     & 4.31$\pm$0.11 & 0.500$\pm$0.221 & 0.181$\pm$0.063 &   5210$\pm$66      & 4.37$\pm$0.11 & 0.555$\pm$0.190 & 0.069$\pm$0.054\\
\citet{melendez14} & 5270$\pm$35     & 4.37$\pm$0.08 & 0.755$\pm$0.133 & 0.174$\pm$0.044 &   5255$\pm$24      & 4.44$\pm$0.06 & 0.767$\pm$0.105 & 0.070$\pm$0.031\\
\citet{morel14}    & 5265$\pm$31     & 4.35$\pm$0.09 & 0.795$\pm$0.102 & 0.197$\pm$0.031 &   5275$\pm$31      & 4.45$\pm$0.08 & 0.870$\pm$0.089 & 0.089$\pm$0.032\\
\citet{reddy03}    & 5320$\pm$38     & 4.51$\pm$0.11 & 0.900$\pm$0.062 & 0.218$\pm$0.036 &   5295$\pm$29      & 4.52$\pm$0.09 & 0.958$\pm$0.046 & 0.092$\pm$0.027\\
\citet{tsantaki19} & 5190$\pm$64     & 4.26$\pm$0.09 & 0.590$\pm$0.149 & 0.163$\pm$0.048 &   5140$\pm$81      & 4.35$\pm$0.08 & 0.485$\pm$0.198 & 0.050$\pm$0.049\\
\hline\hline
\end{tabular}
\end{table*}

\subsection{Stellar abundances}\label{sect_stellar_abundances}
We proceed for the abundance analysis with the extensive line list of \citet{melendez14} because the lines of some important elements (e.g., Mg) in \citet{biazzo12} are not covered by our observations. Hyperfine structure was taken into account for Sc, V, Mn, Co and Cu using atomic data from the Kurucz database\footnote{Available at \url{http://kurucz.harvard.edu/linelists.html}}, while the Eu data were taken from \citet{ivans06}. A classical curve-of-growth analysis making use of the EWs was performed for most species. However, the determination of some abundances relied on spectral synthesis. The oxygen abundance was based on \ion{[O}{i]} $\lambda$630.0, while the C abundance was also estimated from the C$_2$ lines at 508.6 and 513.5 nm. See \citet{morel14} for further details on the modeling of the \ion{[O}{i]} and C$_2$ features. Finally, the Eu abundance was based on a synthesis of a number of \ion{Eu}{ii} lines \citep[for details, see][]{2020A&A...644A..19W}. For \object{K2-138}, $v \sin i$ = 2.5 and a macroturbulence of 1.9 km s$^{-1}$ were assumed based on the analysis reported in \citet{2019A&A...631A..90L}. An attempt was made to model \ion{Li}{i} $\lambda$670.8. The line is not detected in \object{K2-138}, but the Li abundance appears to be much lower than solar.

The abundances are provided in Table \ref{tab_spectroscopic_abundances}. The random uncertainties were estimated following \citet{morel18}. For the spectral synthesis, additional sources of errors (e.g., continuum placement) were taken into account \citep[see][]{morel14}. The O abundance is based on a single line that is weak (EW $<$ 10 m\AA) and blended with a Ni line. It is therefore uncertain. The same is true for the Mg abundance that is based on three strong lines exhibiting quite a large line-to-line scatter ($\sim$0.05 dex).

The impact of lowering $T_{\rm eff}$ by 50 K (see Sect.~\ref{sect_stellar_parameters}) is also given in Table \ref{tab_spectroscopic_abundances}. The Sc, Ti and Cr abundances were derived from both neutral and singly ionised species. Ionisation balance is fulfilled within the uncertainties in all cases assuming the default parameters. However, it can be noted that the agreement systematically degrades for the cooler $T_{\rm eff}$ scale.

\begin{table}[h!]
\caption{Abundance results for \object{K2-138}. The last column shows the impact of lowering $T_{\rm eff}$ by 50 K (see Sect.~\ref{sect_stellar_parameters}), while keeping $\log g$ and $\xi$ unchanged.}
\label{tab_spectroscopic_abundances} 
\centering
\begin{tabular}{l|cc}
\hline\hline
Abundance ratio  & Default $T_{\rm eff}$ scale  & Cooler $T_{\rm eff}$ scale \\
\hline
$[$Fe/H$]$            &  +0.08$\pm$0.05 (42+4) &  +0.01\\
\hline
$[$\ion{C}{i}/Fe$]$   & --0.04$\pm$0.08 (3)    &  +0.03\\
$[$C$_2$/Fe$]$        & --0.07$\pm$0.09 (2)    &  --0.01 \\    
$[$\ion{O}{i}/Fe$]$          &  +0.03$\pm$0.10 (1)    &  --0.01\\
$[$\ion{Na}{i}/Fe$]$         &  +0.02$\pm$0.06 (3)    &  --0.04\\
$[$\ion{Mg}{i}/Fe$]$         &  --0.06$\pm$0.08 (3)   &   --0.05\\
$[$\ion{Al}{i}/Fe$]$         &  +0.01$\pm$0.05 (2)    &  --0.04\\
$[$\ion{Si}{i}/Fe$]$         &  +0.01$\pm$0.04 (10)   &  +0.00\\
$[$\ion{Ca}{i}/Fe$]$         &  +0.04$\pm$0.06 (3)    &  --0.05\\
$[$\ion{Sc}{i}/Fe$]$         &  --0.03$\pm$0.10 (4)   &   --0.06\\
$[$\ion{Sc}{ii}/Fe$]$        &  --0.01$\pm$0.05 (5)   &   --0.01\\
$[$\ion{Ti}{i}/Fe$]$         &  +0.01$\pm$0.08 (14)   &  --0.07\\
$[$\ion{Ti}{ii}/Fe$]$        &  +0.01$\pm$0.06 (10)   &  +0.00\\
$[$\ion{V}{i}/Fe$]$          &  +0.03$\pm$0.08 (5)    &  --0.07\\
$[$\ion{Cr}{i}/Fe$]$         &  +0.03$\pm$0.05 (7)    &  --0.04\\
$[$\ion{Cr}{ii}/Fe$]$        &  +0.08$\pm$0.04 (4)    &  +0.01\\
$[$\ion{Mn}{i}/Fe$]$         &  +0.04$\pm$0.07 (5)    &  --0.05\\
$[$\ion{Co}{i}/Fe$]$         &  +0.00$\pm$0.06 (7)    &  --0.03\\
$[$\ion{Ni}{i}/Fe$]$         &  +0.00$\pm$0.04 (14)   &  --0.02\\
$[$\ion{Cu}{i}/Fe$]$         &  --0.02$\pm$0.03 (2)   &   --0.02\\
$[$\ion{Zn}{i}/Fe$]$         &  --0.01$\pm$0.03 (3)   &   +0.00\\
$[$\ion{Sr}{i}/Fe$]$         &  +0.01$\pm$0.09 (1)    &  --0.07\\
$[$\ion{Y}{ii}/Fe$]$         &  +0.02$\pm$0.07 (4)    &  --0.01\\
$[$\ion{Zr}{ii}/Fe$]$        &  +0.06$\pm$0.06 (2)    &  --0.02\\
$[$\ion{Ba}{ii}/Fe$]$        &  +0.02$\pm$0.07 (1)    &  --0.02\\
$[$\ion{Ce}{ii}/Fe$]$        &  +0.01$\pm$0.08 (5)    &  --0.02\\
$[$\ion{Nd}{ii}/Fe$]$        &  +0.07$\pm$0.05 (3)    &  --0.02\\
$[$\ion{Eu}{ii}/Fe$]$        &  +0.04$\pm$0.08 (3)    &  --0.02\\
\hline
$[$\ion{C}{i}/\ion{O}{i}$]$         & --0.07$\pm$0.13        &  +0.04\\
$[$C$_2$/\ion{O}{i}$]$       & --0.10$\pm$0.12        &  +0.00\\
$[$\ion{Mg}{i}/\ion{Si}{i}$]$       & --0.07$\pm$0.08        &  --0.05\\
\hline\hline
\end{tabular}
\tablefoot{The number in brackets gives the number of lines the abundance is based on. For iron, the number of \ion{Fe}{i} and \ion{Fe}{ii} lines is given.}
\end{table}

\section{\texttt{PASTIS} analysis} \label{sect:pastis}
The joint analysis of the HARPS radial velocities, \textit{K2} light curve and spectral energy distribution (SED) was made using the Bayesian software \texttt{PASTIS} \citep{2014MNRAS.441..983D}. Improvements with respect to our previous analysis in  \citet{2019A&A...631A..90L} are (1) the radial velocities were nightly binned to average out the correlated high-frequency noise resulting from granulation and instrumental calibrations, (2) the new stellar parameters, as derived in section \ref{sect_stellar_parameters}, were used as priors. We ran two sets of analysis with the adopted $T_{\rm eff}$ and lowered by 50 K, as the latter cannot be ruled out, as reported in section \ref{sect_stellar_parameters}.

The magnitudes used to construct the SED were taken from the American Association of Variable Star Observers Photometric All-Sky Survey \citep{2015AAS...22533616H} archive in the optical, from the Two-Micron All-Sky Survey \citep{2014AJ....148...81M} and the Wide-field Infrared Survey Explorer \citep{2014yCat.2328....0C} archives in the near-infrared. The SED was modelled with the BTSettl stellar atmospheric models \citep{2012RSPTA.370.2765A}. The radial velocities were modelled with keplerian orbit models for the planetary contribution and with a gaussian process regression for the correlated noise induced by the activity. For the latter, the following quasi-periodic kernel was used:

\begin{equation}
\begin{split}
    k(t_i, t_j) = A^2 \exp \left[ - \frac{1}{2} \left( \frac{t_i - t_j}{\lambda_1} \right)^2 - \frac{2}{\lambda_2^2} \sin^2 \left( \frac{\pi \left| t_i - t_j \right|}{P_{\rm rot}} \right) \right] \\ + \delta_{ij} \sqrt{\sigma_i^2 + \sigma_J^2}
\end{split}
\end{equation} where A can be identified to the radial velocity modulation amplitude, P$_{\rm rot}$ to the
stellar rotation period, $\lambda_1$ to the correlation decay timescale of the active regions, $\lambda_2$ to the relative contribution between the periodic and the decaying components, and $\sigma_J$ to the radial velocity jitter. To model the photometry, we used the JKT Eclipsing Binary Orbit Program \citep{2008MNRAS.386.1644S} with an oversampling factor of 30 to account for the long integration time of \textit{Kepler} \citep{2010MNRAS.408.1758K}. The star was modelled with the PARSEC evolution tracks \citep{2012MNRAS.427..127B}, taking into account the asterodensity profiling \citep{2014MNRAS.440.2164K}, and with the limb darkening coefficients taken from \citet{2011A&A...529A..75C}.

We ran 80 Markov chain Monte Carlo (MCMC) with $10^6$ iterations for the two different effective temperatures to explore the posterior distributions of the parameters. The convergence was assessed with a Kolmogorov-Smirnov test \citep{10.2307/1391067}. The burn-in phase was then removed \citep{2014MNRAS.441..983D} and the remaining iterations of the different chains having converged were merged. Both analysis, with $T_{\rm eff}$ and $T_{\rm eff}$ lowered by 50 K, converged towards the same distributions, and in particular the same median effective temperature. Therefore we only report the posteriors for the analysis based on $T_{\rm eff} = 5275$ K, along with the priors used. These are shown in Table \ref{MCMCprior}.

The parameters obtained are fully compatible with that of \citet{2019A&A...631A..90L}. In particular, we find masses of $2.80^{+0.94}_{-0.96} ~\Mearth$, $5.95^{+1.17}_{-1.12} ~\Mearth$, $7.20\pm1.40 ~\Mearth$, $11.28^{+2.78}_{-2.72} ~\Mearth$ respectively for planets b, c, d, and e, giving a precision of $34\%$, $20\%$, $19\%$, and $25\%$. For planets f and g, the median values on the masses are respectively $2.43^{+3.05}_{-1.75} ~\Mearth$ and $2.45^{+2.92}_{-1.74} ~\Mearth$, giving a significance of $1.4 ~\sigma$ for both planets. For planet g, the non detection is not surprising given the relatively long orbital period, for a planet with a radius compatible with a low-density planet. Conversely, for planet f, we cannot exclude an absorption of the signal by the gaussian process given its orbital period is half the stellar rotation period. Further discussion on the constraints and upper limits of the planetary masses can be found in \citet{2019A&A...631A..90L}. The parameters of the planets were then used as input for the planets modelling described in the following section.


\section{Composition analysis} \label{sect:methodology}

\subsection{Interior-atmosphere model} \label{sect:interior}

We used the internal structure model initially developed by \citet{2017ApJ...850...93B} and \citet{mousis20}, and recently updated by \cite{2021arXiv210108172A} for the study of their internal composition. The model can accommodate a surface water layer. To consider the effect of the stellar irradiation on this layer, we include a water-rich atmosphere on top of the high-pressure water layer or the mantle by coupling the interior to an atmosphere model. The atmospheric model computes the temperature at the bottom of the atmosphere, which is the boundary condition for the interior model. As a result, our current atmosphere-interior model allows us to assess in detail how well a close-in planet, as the ones we analyze in Sect. \ref{sect:results}, can support a water-rich layer either in liquid, vapour or supercritical state depending on the surface temperature.

Our atmosphere-interior model takes into account the irradiation received by the planet and calculates the surface temperature assuming a water-rich atmosphere on top of a high-pressure water layer or a mantle. Therefore, in Sect. \ref{sect:results}, we use the terms volatile mass fraction and water mass fraction interchangeably. The planets in the multiplanetary systems we analyse are highly irradiated, with irradiation temperatures ranging from approximately 1300 K to 500 K (see Table \ref{output_mcmc}). Depending on the corresponding surface conditions, if water is present, it can be in vapour or supercritical state.

The input variables of the interior structure model are the total planetary mass, the core mass fraction (CMF) and the water mass fraction (WMF), while the model outputs the total planetary radius and the Fe/Si mole ratio. In order to explore the parameter space, we performed a complete Bayesian analysis to obtain the probability density distributions of the parameters. This Bayesian analysis was carried out via the implementation of a MCMC algorithm, by adapting the method proposed by \citet{dorn15} to our interior and atmosphere model as described in \citet{2021arXiv210108172A}.  

Initial values of the three input parameters were randomly drawn from their prior distributions, which correspond to a Gaussian distribution for the mass, and uniform distributions for the CMF and the WMF. We establish a maximum WMF in the uniform prior of 80\%, based on the maximum water content found in Solar System bodies \citep{mckay19}. For the atmosphere, we have considered a composition of 99\% water and 1\% carbon dioxide. The atmosphere and the interior are coupled at a pressure of 300 bar. We consider the stellar spectral distribution of a Sun-like star for the calculation of the Bond albedo. The atmospheric mass, thickness, Bond albedo, and temperature at the bottom of the atmosphere are provided by a grid generated with the atmospheric model described in \citet{marcq17} and \cite{pluriel19}.

\subsection{Atmospheric escape}
\label{sect:escape}

Atmospheric mass loss in super-Earths and sub-Neptunes can be produced by thermal or non-thermal escape, with Jeans escape \citep{jeans25}, XUV photoevaporation \citep{owen12} or core-powered mass loss \citep{ginzburg16}. These processes might shape the trend of the volatile mass fraction (water, H/He or a combination of both) in the inner region of multiplanetary systems. An estimate of the mass loss rates of different species can discriminate between two possible interior compositions. In our solar system, Jeans' escape efficiently removed lighter gases as H$_2$ and He on telluric planets, leaving heavier molecules. For the planets in the K2-138 system, we estimate Jeans mass loss rates \citep{aguichine21} by using as input the masses, radii and equilibrium temperatures we obtained as a result of our spectroscopic analysis (Sect. \ref{sect_spectroscopic_analysis}). For the rest of the multiplanetary systems we analyse, we use the parameters provided by the references we mention in Sect. \ref{sect:multipl_data}.

The hydrodynamic escape of H-He is driven by the incident XUV flux from the host star. A star's XUV luminosity $L_{\mathrm{XUV}}$ is usually constant at early stages, called saturation regime (a few tens of Myr), and then evolves as a power-law function of time $L_{\mathrm{XUV}}\propto t^\alpha$, with $\alpha\simeq -1.5$ \citep{sanzforcada11}. Computing the mass loss rate from \citep{owen12}:
\begin{equation}
    \dot{m} = \eta \frac{L_{\mathrm{XUV}} R_b^3}{GM_b (2a_b)^2},
    \label{eq:dotm-xuv}
\end{equation}

where $G$ is the gravitational constant and $\eta=0.1$ is an efficiency factor \citep{owen12}. Following the approach in \cite{aguichine21}, we integrate Equation (\ref{eq:dotm-xuv}) over time assuming that only $L_{\mathrm{XUV}}$ can vary, implying mass and radius do not change significantly, to calculate the total lost mass.

\subsection{Multiplanetary systems parameters}
\label{sect:multipl_data}

In addition to K2-138, we select a sample of multiplanetary systems that host only low-mass planets ($M$ < 20 $M_{\oplus}$), with five or more planets that have masses and radii available. These systems are TOI-178, Kepler-11, Kepler-102 and Kepler-80. For K2-138, we take the planetary mass and radius derived in section \ref{sect:pastis}, and the corrected Fe/Si molar ratio. The latter was estimated as Fe/Si = 0.77$\pm$0.07, using the metallicity and the Mg, Al, Si, Ca and Ni abundances presented in section \ref{sect_stellar_abundances}, following \cite{sotin07} and \cite{2017ApJ...850...93B}.

For the other systems, we performed the same modeling, taking masses, radii and stellar abundances from \cite{Leleu21} for TOI-178; \cite{Lissauer11} and \cite{Brewer16} for Kepler-11; \cite{Marcy14} and \cite{Brewer18} for Kepler-102, and \cite{Macdonald16} and \cite{Macdonald21} for Kepler-80. The Fe/Si mole ratios of these systems are computed similarly to the Fe/Si mole ratio of K2-138 from their respective host stellar abundances.

\begin{table*}[]
\centering
\begin{tabular}{cccccc}
\hline \hline
System & Planet & M [$M_{\oplus}$] & R [$R_{\oplus}$] & $a_{d}$ [AU] & $T_{irr}$ [K] \\
\hline
\multirow{6}{*}{TOI-178} & b & 1.5$^{+0.39}_{-0.44}$ & 1.152$^{+0.073}_{-0.070}$ & 0.026 & 1040 \\
 & c & 4.77$^{+0.55}_{-0.68}$ & 1.669 $^{+0.114}_{-0.099}$ & 0.037 & 873 \\
 & d & 3.01$^{+0.80}_{-1.03}$ & 2.572$^{+0.075}_{-0.078}$ & 0.059 & 691 \\
 & e & 3.86$^{+1.25}_{-0.94}$ & 2.207$^{+0.088}_{-0.090}$ & 0.078 & 600 \\
 & f & 7.72$^{+1.67}_{-1.52}$ & 2.287$^{+0.108}_{-0.110}$ & 0.104 & 521 \\
 & g & 3.94$^{+1.31}_{-1.62}$ & 2.87$^{+0.14}_{-0.13}$ & 0.128 & 471 \\ \hline
\multirow{5}{*}{Kepler-11} & b & 4.3$^{+2.2}_{-2.0}$ & 1.97$\pm$0.19 & 0.091 & 953 \\
 & c & 13.5$^{+4.8}_{-6.1}$ & 3.15$\pm$0.30 & 0.106 & 883 \\
 & d & 6.1$^{+3.1}_{-1.7}$ & 3.43$\pm$0.32 & 0.159 & 721 \\
 & e & 8.4$^{+2.5}_{-1.9}$ & 4.52$\pm$0.43 & 0.194 & 653 \\
 & f & 2.3$^{+2.2}_{-1.2}$ & 2.61$\pm$0.25 & 0.250 & 575 \\ \hline
\multirow{5}{*}{Kepler-102} & b & 0.41$\pm$1.6 & 0.47$\pm$0.02 & 0.055 & 868 \\
 & c & -1.58$\pm$2.0 & 0.58$\pm$0.02 & 0.067 & 786 \\
 & d & 3.80$\pm$1.8 & 1.18$\pm$0.04 & 0.086 & 597 \\
 & e & 8.93$\pm$2.0 & 2.22$\pm$0.07 & 0.117 & 694 \\
 & f & 0.62$\pm$3.3 & 0.88$\pm$0.03 & 0.165 & 501 \\ \hline
\multirow{5}{*}{Kepler-80} & d & 5.95$^{+0.65}_{-0.60}$ & 1.309$^{+0.036}_{-0.032}$ & 0.033 & 990 \\
 & e & 2.97$^{+0.76}_{-0.65}$ & 1.330$^{+0.039}_{-0.038}$ & 0.044 & 863 \\
 & b & 3.50$^{+0.63}_{-0.57}$ & 2.367$^{+0.055}_{-0.052}$ & 0.058  & 750 \\
 & c & 3.49$^{+0.63}_{-0.57}$ & 2.507$^{+0.061}_{-0.058}$ & 0.071 & 679 \\
 & g & 0.065$^{+0.044}_{-0.038}$ & 1.05$^{+0.22}_{-0.24}$ & 0.094 & 588 \\
 \hline
\end{tabular}
\caption{Masses, radii, semi-major axis and irradiation temperature for the multiplanetary systems TOI-178, Kepler-11, Kepler-102 and Kepler-80. References can be found in Sect. \ref{sect:multipl_data}. }
\label{tab:my-table}
\end{table*}

\section{Compositional trends in multiplanetary systems} \label{sect:results}

Table \ref{tab:multiplanets} shows the retrieved CMF and WMF and their one-dimensional 1$\sigma$ uncertainties as a result of our Bayesian analysis, as well as their atmospheric mass loss estimates. To assess how compatible a water-rich composition is with the data, we also show the difference between the observational mean and the retrieved mean, which is calculated as $d_{obs-ret}$ =  max$\left\lbrace | R_{data}-R | , | M_{data}-M | \right\rbrace $. If $d_{obs-ret}$ is below 1$\sigma$, the retrieved mass and radius agree within the 1$\sigma$ confidence intervals with the observed mass and radius, meaning that the density of a planet is compatible with a volatile layer dominated by water. A high $d_{obs-ret}$  (> 1 $\sigma$), and a high WMF in our model simultaneously, indicate that a water-dominated atmosphere is not inflated enough to account for the low density of the planet, pointing to an atmosphere with more volatile gases, which are probably H and He. Table \ref{output_mcmc} shows the irradiation temperatures and the retrieved atmospheric parameters of the planets whose density is compatible with the presence of a volatile layer dominated by water.

\subsection{K2-138} 

\begin{table*}[h]
\centering
\begin{tabular}{cccccccc}
\hline \hline
System & Planet & CMF & WMF & $d_{obs-ret}$ & $\Delta M_{H2}$ [$M_{\oplus}$] & $\Delta M_{H2O}$ [$M_{\oplus}$] & $\Delta M_{XUV}$ [$M_{\oplus}$] \\ \hline

\multirow{6}{*}{K2-138} & b & 0.27$\pm$0.02 & 0.000$_{-0.000}^{+0.007}$ & 1.5 $\sigma$ & 0.132 & < 0.01 & 0.40 \\
 & c & 0.23$\pm$0.02 & 0.13$\pm$0.04 & \textless 1 $\sigma$ & < 0.01 & < 0.01 & < 0.01 \\ 
 & d & 0.22$\pm$0.03 & 0.17$\pm$0.05 & \textless 1 $\sigma$ & < 0.01 & < 0.01 & < 0.01 \\
 & e & 0.11$\pm$0.02 & 0.57$\pm$0.08 & \textless 1 $\sigma$ & < 0.01 & < 0.01 & < 0.01 \\
 & f & 0.11$\pm$0.02 & 0.60$\pm$0.07  & \textless 1 $\sigma$ & < 0.01 & < 0.01 & < 0.01 \\
 & g & 0.12$\pm$0.05 & 0.55$\pm$0.18 &  1.3 $\sigma$ & < 0.01 & < 0.01 & < 0.01 \\ \hline
\multirow{6}{*}{TOI-178} & b & 0.21$\pm$0.30 & 0 & \textless 1 $\sigma$ & 0.83 & < 0.01 & 0.45 \\
 & c & 0.30$\pm$0.02 & 0.02$^{+0.04}_{-0.02}$ & \textless 1 $\sigma$ & < 0.01 & < 0.01 & 0.21 \\
 & d & 0.10$\pm$0.01 & 0.69$\pm$0.05 & 1.3 $\sigma$ & 0.16 & < 0.01 & 0.48 \\
 & e & 0.18$\pm$0.02 & 0.40$\pm$0.06 & \textless 1 $\sigma$ & < 0.01 & < 0.01 & 0.13 \\
 & f & 0.22$\pm$0.03 & 0.28$\pm$0.10 & \textless 1 $\sigma$ & < 0.01 & < 0.01 & 0.04 \\
 & g & 0.10$\pm$0.01 & 0.58$\pm$0.16 & 3.0 $\sigma$ & < 0.01 & < 0.01 & 0.11 \\ \hline
\multirow{5}{*}{Kepler-11} & b & 0.20$\pm$0.04 & 0.27$\pm$0.10 & \textless 1 $\sigma$ & < 0.01 & < 0.01 & 0.10 \\
 & c & 0.18$\pm$0.01 & 0.33$\pm$0.04 & 1.7 $\sigma$ & < 0.01 & < 0.01 & 0.10 \\
 & d & 0.10$\pm$0.02 & 0.65$\pm$0.05 & 2.4 $\sigma$ & < 0.01 & < 0.01 & 0.13 \\
 & e & 0.12$\pm$0.01 & 0.55$\pm$0.04 & 4.4 $\sigma$ & < 0.01 & < 0.01 & 0.14 \\
 & f & 0.14$\pm$0.06 & 0.47$\pm$0.10 & 1.9 $\sigma$ & 0.56 & < 0.01 & 0.06 \\ \hline
\multirow{5}{*}{Kepler-102} & b & 0.91$^{+0.09}_{-0.16}$ & 0 & \textless 1 $\sigma$ & 0.13 & < 0.01 & 0.03 \\
 & c & 0.95$^{+0.05}_{-0.30}$ & 0 & \textless 1 $\sigma$ & 0.10 & < 0.01 & 0.03 \\
 & d & 0.80$\pm$0.14 & 0 & \textless 1 $\sigma$ & < 0.01 & < 0.01 & 0.03 \\
 & e & 0.22$\pm$0.02 & 0.17$\pm$0.07 & \textless 1 $\sigma$ & 0.01 & < 0.01 & 0.03 \\
 & f & 0.27$\pm$0.09 & 0.04$\pm$0.04 & \textless 1 $\sigma$ & 0.02 & < 0.01 & 0.01 \\ \hline
 \multirow{5}{*}{Kepler-80} & d & 0.97 $^{+0.03}_{-0.05}$ & 0 & \textless 1 $\sigma$ & < 0.01 & < 0.01 & 0.35 \\
 & e & 0.43$\pm$0.18 & 0 & \textless 1 $\sigma$ & < 0.01 & < 0.01 & 0.29 \\
 & b & 0.13$\pm$0.02 & 0.58$\pm$0.07 & \textless 1 $\sigma$ & < 0.01 & < 0.01 & 0.11 \\
 & c & 0.09$\pm$0.01 & 0.70$\pm$0.04 & \textless 1 $\sigma$ & < 0.01 & < 0.01 & 0.13 \\
 & g & 0.31$\pm$0.02 & < 1.5 $\times \ 10^{-3}$ & \textless 1 $\sigma$ & 140 & 3.23 & 0.60 \\ \hline

\end{tabular}
\caption{Retrieved core mass fraction (CMF) and water mass fraction (WMF) of planets in the multiplanetary systems K2-138, TOI-178, Kepler-11, Kepler-102 and Kepler-80, with our interior-atmosphere model. A low $d_{obs-ret}$ indicates that the assumption of a water-dominated atmosphere is adequate for a particular planet (see text). $\Delta M_{H2}$, $\Delta M_{H2O}$ and $\Delta M_{XUV}$ correspond to the maximum estimate of atmospheric escape mass loss due to H$_{2}$, H$_{2}$O Jeans escape and XUV photoevaporation, respectively.}
\label{tab:multiplanets}
\end{table*}

\begin{table*}[h]
\centering

\begin{tabular}{ccccc}
\hline \hline
Planet &  $T_{irr}$ [K]  & $T_{300}$ [K] & $z_{atm}$ [km] & $A_{B}$ \\ \hline
K2-138 b & 1291 & 4110$\pm$44 & 932$\pm$151 & 0.213$\pm$0.001 \\
K2-138 c & 1125 & 3900$\pm$23 & 711$\pm$103 & 0.214$\pm$0.002 \\
K2-138 d & 978 & 3614$\pm$56 & 635$\pm$84 & 0.218$\pm$0.002 \\
K2-138 e & 850 & 3383$\pm$39 & 673$\pm$90 & 0.231$\pm$0.001 \\
K2-138 f & 735 & 3396$\pm$116 & 1483$\pm$546 & 0.260$\pm$0.004 \\
TOI-178 c & 873 & 3344$\pm$33 & 500$\pm$60 & 0.226$\pm$0.001 \\
TOI-178 d & 691 & 3254$\pm$45 & 1181$\pm$224 & 0.264$\pm$0.004 \\
TOI-178 e & 600 & 2930$\pm$31 & 690.7$\pm$133 & 0.225$\pm$0.018 \\
TOI-178 f & 521 & 2610$\pm$23 & 368$\pm$60 & 0.298$\pm$0.007 \\
Kepler-11 b & 953 & 3697$\pm$133 & 840$\pm$313 & 0.221$\pm$0.005 \\
Kepler-102 e & 694 & 2947$\pm$29 & 360$\pm$55 & 0.243$\pm$0.004 \\
Kepler-102 f & 501 & 2784$\pm$102 & 837$\pm$290 & 0.347$\pm$0.013 \\
Kepler-80 b & 750 & 3344$\pm$33 & 1133$\pm$148 & 0.253$\pm$0.002 \\
Kepler-80 c & 679 & 3219$\pm$29 & 1128$\pm$114 & 0.266$\pm$0.003 \\  \hline

\end{tabular}

\caption{Atmospheric parameters retrieved for the planets whose composition can accommodate a water-dominated atmosphere (see text). These parameters are the equilibrium temperature assuming a null albedo ($T_{irr}$), the atmospheric temperature at 300 bar ($T_{300}$), the thickness of the atmosphere from 300 bar to 20 mbar ($z_{atm}$), and the planetary Bond albedo ($A_{B}$).}

\label{output_mcmc}
\end{table*}

  \begin{figure}
   \centering
   \includegraphics[width=\hsize]{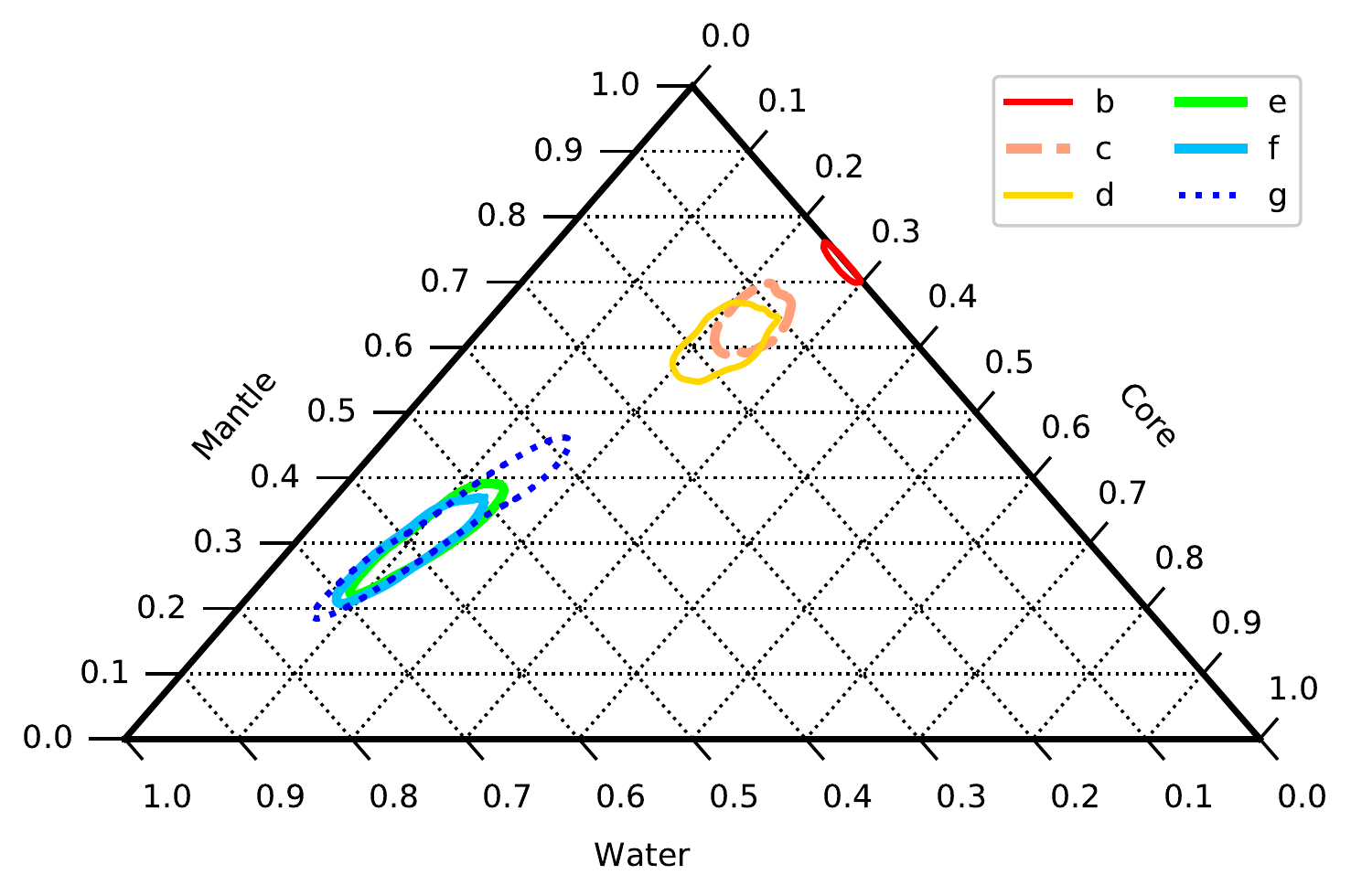}
      \caption{1-$\sigma$ confidence regions derived from the 2D posterior distributions of the CMF and WMF obtained with the planetary interior Bayesian analysis. Axes indicate the core mass fraction (CMF), water mass fraction (WMF) and the mantle mass fraction (MMF). The latter is defined as MMF = 1 - (CMF+WMF).}
         \label{ternary}
   \end{figure}
   
        \begin{figure}
   \centering
   \includegraphics[width=\hsize]{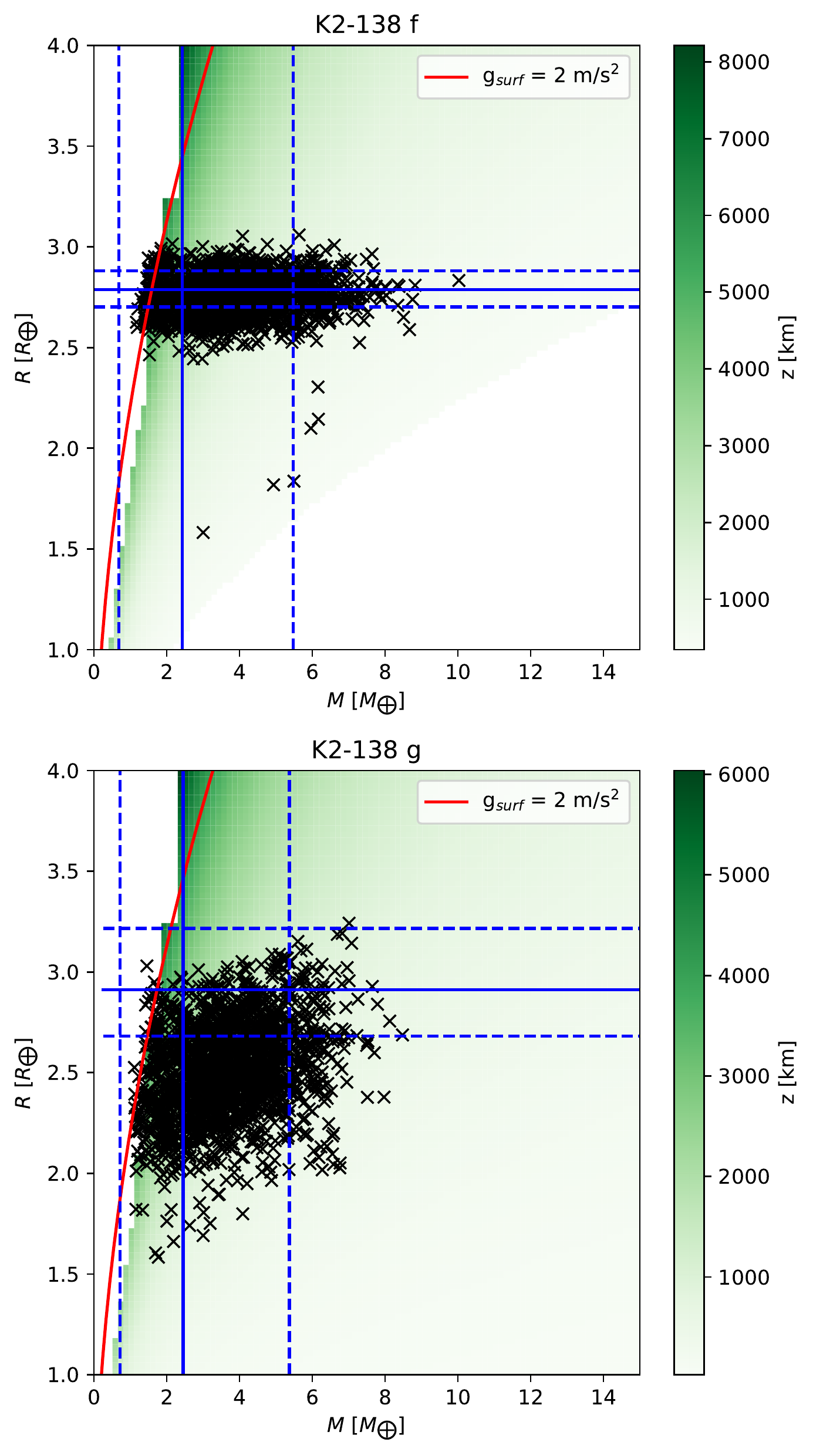}
      \caption{Total mass and radius of K2-138 f (upper panel) and K2-138 g (lower panel) from the different realisations of the MCMC (black crosses). The solid blue lines show the mass and radius measurements from \texttt{PASTIS}, and the dashed lines give the related uncertainties. The red line indicates the limit below which the planet cannot maintain an atmosphere.}
         \label{fig:K2-138f}
   \end{figure}

Figure~\ref{ternary} displays the 1$\sigma$ confidence intervals derived from the 2D distributions of the WMF and CMF of the K2-138 in a ternary diagram. We can see that the confidence regions are aligned along a line almost parallel to the lines where the CMF is constant. This alignment is due to the the constraint on the Fe/Si mole ratio we have considered within the whole planetary system: the confidence regions are spread over the Fe/Si-isolines whose constant values range from Fe/Si = 0.70 to 0.84 \citep[see][their Figure 4]{2017ApJ...850...93B}.

For K2-138 b, the results set an upper limit of 0.7\% in the WMF, which means that this planet is unlikely to have a significant amount of volatiles, including water. The retrieved planetary radius is 1.538 $R_{\oplus}$, which is 1.5$\sigma$ larger than the measured radius from the analysis in section \ref{sect:pastis}. This is due to the extended atmosphere necessary to produce temperature and pressure conditions to hold supercritical water on the surface ($P_{surf} > 300 $ bar). If we assume a mass of 2.80 $M_{\oplus}$ and a CMF of 0.27, a vapour atmosphere with a maximum surface pressure of 300 bar would yield a WMF of 0.01\% (WMF of Earth is 0.05\%) and a radius of 1.461 $R_{\oplus}$, which is well within the 1$\sigma$ confidence interval of the observed value. Therefore, we can conclude that K2-138 b is a volatile-poor planet, that might present a secondary atmosphere with a low surface pressure ($P_{surf} \leq 300 $ bar) or no atmosphere (WMF = 0). In addition, it is the planet with the highest CMF in the system, showing that planets in this system are likely to have less massive cores than Earth (CMF = 0.325) and the other terrestrial planets in the Solar System. 

The atmospheric model also establishes a minimum surface gravity of 2 m/s to retain an atmosphere. Unlike planets b, c, d and e, in which the 1-$\sigma$ intervals on the masses exclude such low surface gravity, this is not the case for planets f and g. For planet f, a lower limit on the surface gravity of the planet can be translated to a lower limit on the mass. If it is below this limit, the gravity at the surface would not be enough to retain an atmosphere. For planet f, with a total radius of 2.762 $R_{\oplus}$ and a CMF of 0.11, this limit would be approximately 2 $M_{\oplus}$. This minimum mass value to retain its atmosphere is above the lower limit of the total mass set by its 1 $\sigma$ uncertainties, as can be seen in Figure \ref{fig:K2-138f}, upper panel. Furthermore, planet f is the most water-rich in the K2-138 planetary system, with an upper limit of 66\% in the WMF, which is close to the 77\% maximum limit on the water content derived from measurements on cometary compositions. Similarly, planet g also presents a lower limit on the mass of the bulk of the planet of $\backsim$ 2 $M_{\oplus}$ (see Figure \ref{fig:K2-138f}, lower panel). Its retrieved planetary radius is significantly lower than the observational value, with a difference of 1.3 $\sigma$. Therefore, the atmosphere of K2-138 g is significantly more extended than an atmosphere dominated by water vapour under the same irradiance conditions. This increase in atmospheric thickness is probably due to an atmosphere rich in H and He. K2-138 g could have up to 5\% of volatile mass fraction assuming a H/He atmosphere \citep[see Fig. 1 in][]{lopez_fortney14}.

A rough estimate of Jeans mass loss rates for K2-138 b yields $6\times 10^{-7}$ $\Mearth.\mathrm{Gyr}^{-1}$ for Jeans escape of H$_2$, and $5\times 10^{-84}$ $\Mearth.\mathrm{Gyr}^{-1}$ for Jeans escape of H$_2$O. For comparison, in the case of Earth the absence of H$_2$ is due to an exobase (altitude at which particles escape) temperature much higher than the equilibrium temperature \citep{hedin83}. An exobase temperature 2 times higher than the equilibrium temperature gives a mass-loss rate of $4\times 10^{-2}$ $\Mearth.\mathrm{Gyr}^{-1}$. In that case, an envelope of 1--10\% of H-He mixture could be efficiently removed, leaving only heavier species such as H$_2$O. In the case of hydrodynamic escape, we obtain a mass loss rate of 2 $\Mearth$.$\mathrm{Gyr}^{-1}$ during the saturation regime and $1\times 10^{-2}$ $\Mearth$.$\mathrm{Gyr}^{-1}$ at $t=3$ Gyr. This yields an integrated mass loss of $0.4\Mearth$, or 14\% of planet's b total mass. Comparing this value to the WMF derived for planets c and d from the MCMC in Table \ref{tab:multiplanets}, we conclude that K2-138 b could have formed with a thick envelope of H$_2$O that has been blown away by XUV photoevaporation.

\subsection{TOI-178}

In the TOI-178 system, planets b and c have an increasing WMF with progressing distance from the star, while planets d to g have WMF equal or greater than 30\%. For planets d and g, the the volatile layer is likely to present H/He, which would explain why in our analysis their WMF are in the 60-70\% range in addition to $d_{obs-ret}$ greater than 1$\sigma$. TOI-178 b could have lost up to 0.83 $M_{\oplus}$ of its current mass in H$_{2}$ due to Jeans escape, and up 0.45 $M_{\oplus}$ due to photoevaporation, while TOI-178 c could have lost 0.21 $M_{\oplus}$. In such scenario, TOI-178 b and c original volatile mass fraction would be up to 0.36 and 0.10, respectively compared to their current value.

\subsection{Kepler-11}

For Kepler-11, the WMF of the innermost planet is 0.27$\pm$0.10, which is compatible with a water-dominated envelope. For Kepler-11 c to e, their radius data are 1.7$\sigma$, 2.4$\sigma$ and 4.4$\sigma$ higher than the radius we retrieve with our model, discarding the water-rich envelope hypothesis. The increasing significance level indicates that these planets have an increasing content of H/He with distance from the star. In the case of the outermost planet, Kepler-11 f, the retrieved radius is 1.9$\sigma$ lower than the data, suggesting that this planet presents less H/He than planets c to e. Nonetheless, this could be because of Kepler-11 f not being able to retain a primordial atmosphere due to its low mass (2.3 $^{+2.2}_{-1.2}$ $M_{\oplus}$), compared to the higher masses of the rest of the planets in the system (> 6 $M_{\oplus}$). Furthermore, Kepler-11 f could have lost up to 0.56 $M_{\oplus}$ in H$_{2}$, according to our atmospheric Jeans escape calculation, whereas the other four planets in the system have atmospheric mass losses below 2$\times 10^{-3} \ M_{\oplus}$.

\subsection{Kepler-102}

The densities of the three innermost planets of Kepler-102 suggest that these are dry planets with high CMFs. Their core-to-mantle ratios could be even higher than the CMF we would expect from the Fe and Si stellar abundances of their host star. Therefore, we set the WMF equal to zero in our MCMC Bayesian analysis and let the CMF as the only free parameter. We only take into account the mass and radius as observables. Our modelling shows that Kepler-102 b, c and d are dry Mercury-like planets, with CMF = 0.91$^{+0.09}_{-0.16}$, 0.95$^{+0.05}_{-0.30}$ and 0.80$\pm$0.14, respectively. Their high CMF could be due to mantle evaporation \citep{Cameron85}, impacts \citep{Benz88,Benz07,Asphaug14} or planet formation in the vicinity of the rocklines \citep{Aguichine20,Scora20}. Kepler-102 e presents a WMF of 0.17$\pm$0.07, suggesting that this planet has a more volatile-rich composition than the planets that precede it. The large uncertainties in the mass of Kepler-102 f prevent us from determining whether this is a bare rocky planet with no atmosphere, or if it presents a thin atmosphere with a maximum WMF = 0.08. In addition, Jeans H$_{2}$ atmospheric escape could have removed up to 0.02 M$_{\oplus}$ from Kepler-102 f, yielding an original volatile mass fraction between 0.07 and 0.10.

\subsection{Kepler-80}

Kepler-80 d presents a high CMF, corresponding to a Fe-rich planet, similarly to Kepler-102 b and c. Kepler-80 e is consistent with a dry planet with an Earth-like CMF, whereas Kepler-80 b and c are volatile-dominated planets. Kepler-80 g shows a WMF of up to 0.15\%. Given its low mass M = 0.065$^{+0.044}_{-0.038} \ M_{\oplus}$ \citep{Macdonald21}, planet g could have not retained a H/He atmosphere, making a secondary atmosphere with water and/or CO$_{2}$ the most likely atmospheric composition for this planet. Based on our MCMC interior-atmosphere analysis, this atmosphere could be of less than 300 bar of surface pressure. This scenario is also supported by our estimated Jeans water escape, which is between 3.26 $\times \ 10^{-3} \ M_{\oplus}$ and 3.24 $M_{\oplus}$. Both Jeans escape and XUV photoevaporation could have removed efficiently a H/He envelope. The total atmospheric mass loss and the current mass add up to a planetary mass that is similar to that of Kepler-80 e, b and c. Finally, the radius of Kepler-80 g is 2.7 $\sigma$ higher than the radius of a rocky planet with no atmosphere, which suggests that Kepler-80 g probably has retained a gaseous envelope.

\section{Discussion} \label{sect:discussion}

\begin{figure*}
   \centering
   \includegraphics[width=0.6\textwidth]{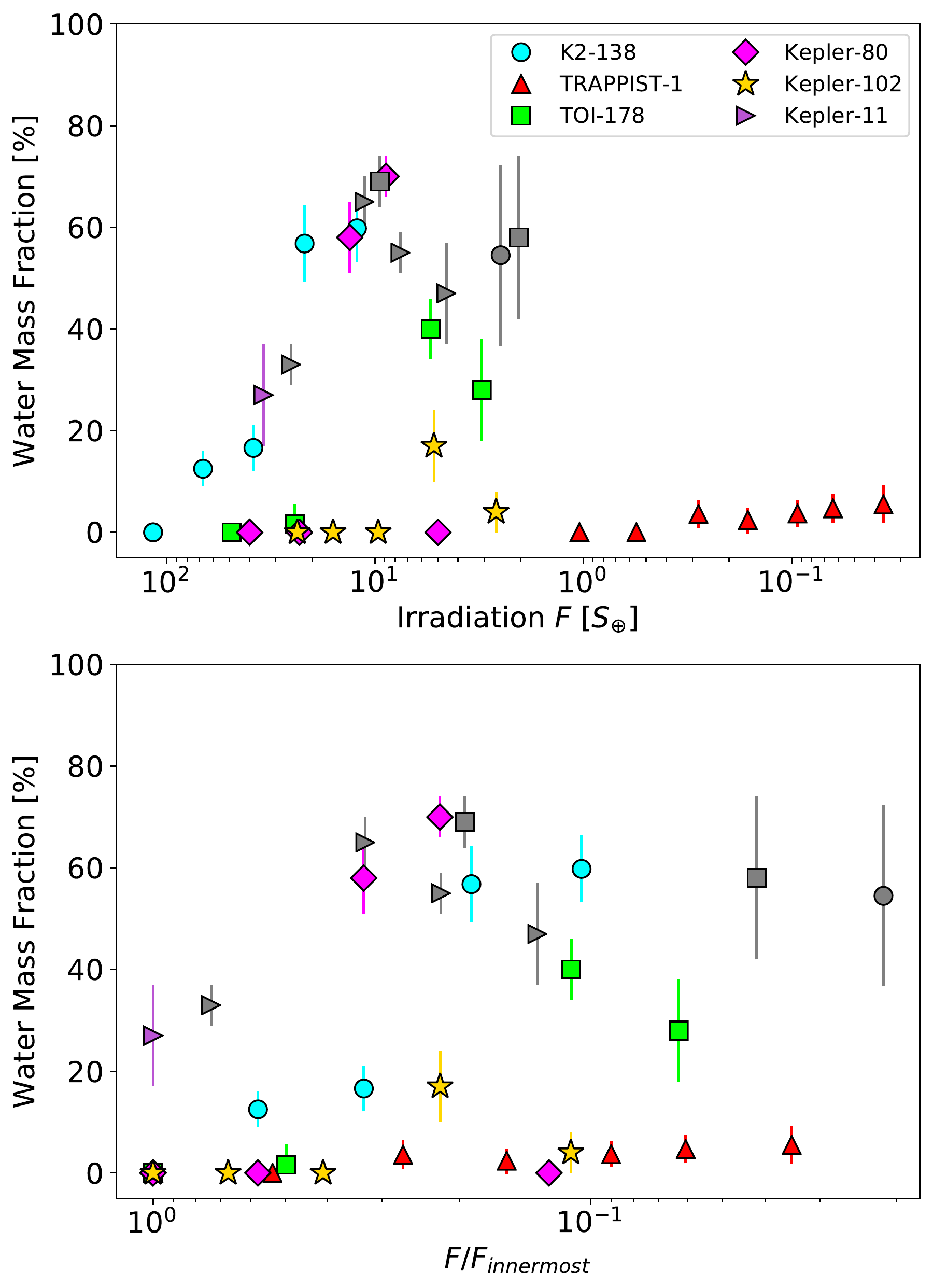}
      \caption{Volatile mass fraction trends of the six multiplanetary systems analysed with our interior-atmosphere model. We show the water mass fraction estimates (see text) as a function of the stellar incident flux or irradiation, $F$, in Earth irradiation units ($S_{\oplus}$ = 1361 $W/m^{2}$) in the upper panel. In the lower panel, the incident flux is normalised with respect to the inner, most irradiated planet in each system, $F_{innermost}$. Planets whose atmospheric composition is likely to be H/He-dominated instead of water-dominated ($d_{obs-ret}$ > 1 $\sigma$) are indicated in grey color. }
         \label{distance}
   \end{figure*}

Figure~\ref{distance} shows the volatile content of the five multiplanetary systems we analysed in this work as a function of the incident flux normalised with the incident flux received by the innermost planet. In addition, we include in Figure~\ref{distance} the WMF of TRAPPIST-1 derived with our interior-atmosphere model by \cite{2021arXiv210108172A} for a homogeneous comparison. Of all systems, K2-138 presents a very clear volatile mass fraction trend: an increasing gradient in water content with distance from the host star for planets b to d, followed by a constant volatile mass fraction for the outer planets (planets e to g). A similar trend is observed in the TRAPPIST-1 system, if one neglects TRAPPIST-1 d presenting a higher volatile mass fraction than its two surrounding inner and outer planets in Fig. ~\ref{distance}. In \cite{2021arXiv210108172A}, the WMF is obtained by assuming a condensed water layer. However, water could be in vapour phase and mixed with CO$_{2}$ in a CO$_{2}$- dominated atmosphere, lowering the overall volatile mass fraction of TRAPPIST-1 d. In that case, the TRAPPIST-1 system could potentially show the increase-plus-plateau volatile trend observed in K2-138. Transmission spectroscopy of TRAPPIST-1 d is needed to probe the composition of its atmosphere. The multiplanetary systems TOI-178 and Kepler-11 do not show smooth increases of the water mass fraction with orbital distances, although the inner planets present significantly less volatiles than the outer planets. Finally, Kepler-80 and Kepler-102 could form this trend if it was not because of their outermost planet, which presents a lower volatile mass fraction than the planet that immediately precedes it. In addition, the estimated original volatile mass fraction of Kepler-102 f is well within the uncertainties of the WMF of Kepler-102 e, meaning that the planets e and f could potentially form a plateau in the outer part of the Kepler-102 system with a water mass fraction of 10\%, similarly to TRAPPIST-1.

In the case of the TOI-178 and Kepler-11, it would be necessary to adopt a self-consistent modelling approach that includes the possibility of a H/He-dominated volatile layer to determine whether their volatile mass fraction trend is as clear as that of K2-138 and TRAPPIST-1. For the other multiplanetary systems, which do not present high $d_{obs-ret}$ combined with high water mass fractions in our analysis, the volatile mass fraction would decrease for each individual planet under the assumption of a H/He envelope. Including H/He as part of the envelope would change the value of the volatile mass fraction of each individual planet, but it would not change our conclusion about the global volatile mass fraction trends in each system (i.e the gradient and plateau trend in TRAPPIST-1 and K2-138).  Furthermore, the water-H/He degeneracy to which volatile-rich planets are subject to can only be broken with atmospheric characterization data, such as transmission spectroscopy and phase curves. In many cases, the volatile envelope of sub-Neptunes might not be dominated by either water or H/He, but it could be a mixture of both. This is supported by transmission spectroscopy of the sub-Neptune K2-18 b \citep{Tsiaras19,Benneke19,Madhusudhan20}, where water is detected, although its current trace species could be compatible with a H$_{2}$-rich atmosphere \citep{Yu21}. Additionally, meteorite outgassing experiments show that a significant fraction of H/He could be sustained in a water-dominated secondary atmosphere \citep{Thompson21}.

The significant difference in volatile mass fraction between the inner planets and the outer planets of these multiplanetary systems indicates that these planets might have undergone similar formation and evolution histories. The gradient-plus-plateau trend could potentially result from the combination of planetary formation in ice-rich regions of the protoplanetary disk, atmospheric loss, and inward migration. The outer volatile-rich planets could have formed beyond the ice line prior to migration, where ice-rich solids are expected to form \citep{Mousis21}, producing planets with high volatile contents. In the systems whose planets present water mass fractions lower than 10\%, volatiles could have been simply delivered by building blocks made of chondritic minerals bearing this amount of water \citep{Daswani21}. In those conditions, the radial drift of icy planetesimals from beyond the snowline is not required. In the case of K2-138, the three-body Laplace resonances are a sign of an inner planetary migration \citep{2007ApJ...654.1110T, 2017MNRAS.470.1750I, 2017A&A...602A.101R}. For three systems, we found that their outermost planets (Kepler-11 f, Kepler-102 f and Kepler-80 g) have lower volatile mass fractions than the planets before them in the system. This could be due to their lower masses compared to the other planets in their systems, since they are not massive enough to have a surface gravity that would help them retain their atmospheres. In addition, these three low-mass, low-WMF planets could have formed further away from the water ice line than the water-rich planets in their systems, having less water-rich material available during accretion than those planets that formed in the vicinity of the water ice line.

Contrasting with K2-138, the water mass fraction of the outer planets found in the planets of the TRAPPIST-1 and Kepler-102 systems are compatible with 10\% \citep{agol21,2021arXiv210108172A}, a value found in agreement with the water content of many asteroids of the Main Belt \citep{Vernazza15}. This similarity suggests that the building blocks of the outer planets of these systems could have agglomerated from a mixture of ice grains coming from the snowline and anhydrous silicates formed at closer distances from the host star, following the classical formation scenarios invoked for the Main Belt \citep{2002aste.book..235R}. In that case, this implies that the migration distances of the planets in TRAPPIST-1 and Kepler-102 would have been more restricted than those of the water-rich planets in the K2-138, TOI-178 and Kepler-11 systems.

We have considered the Fe/Si mole ratio as an observable of our MCMC Bayesian analysis in addition to the planetary masses and radii. Even though the Fe/Si derived from stellar abundances and that obtained from rocky planet densities could depart from a 1:1 relationship \citep{Plotnykov20,Adibekyan21}, considering the Fe/Si mole ratio contributes to reducing the degeneracy between the rock+mantle layers and the volatile layer \citep{dorn15,Dorn17,2017ApJ...850...93B}. Particularly, assuming that the planetary Fe/Si mole ratio is similar to the Fe/Si ratio of the host star improves the determination of the CMF, but does not necessarily contribute to the determination of the volatile mass fraction in volatile-rich planets \citep{Otegi20}. This is the case of the TRAPPIST-1 system, where the inclusion of the Fe/Si mole ratio as an observable in the MCMC Bayesian analysis refines the determination of the surface pressure for the inner planets of the system, but slightly reduces the uncertainties of the WMF estimates for the outer planets \citep[see Tables 3 and 4 in][]{2021arXiv210108172A}. Therefore, considering the Fe/Si mole ratio does not affect the volatile general trend of the planets within a multiplanetary system.

\section{Conclusions} \label{sect:conclusion}

We carried out a homogeneous interior modelling and composition analysis of five multiplanetary systems that have 5 or more low-mass planets ($M < 20 \ M_{\oplus}$), rather than compiled the volatile content estimates of previous works, to eliminate the differences between interior models as a possible bias when comparing the compositional trends between planetary systems. In the case of the TOI-178, Kepler-11, Kepler-102 and Kepler-80 systems, we used previously published mass, radius and stellar abundances data. In the case of the K2-138 system, we completed the previous analysis with an in-depth stellar spectroscopic analysis. We performed a line-by-line differential analysis of K2-138 spectra with respect to $\alpha$ Cen B and the Sun, to derive the most accurate stellar parameters and abundances given the data at hand. These were used for a new complete Bayesian analysis of the radial velocities and photometry acquired on the system. We explored the robustness of the planetary parameters and stellar chemical abundances in our spectroscopic analysis. We concluded that the parameters we derived are fully consistent with the ones obtained by \citet{2019A&A...631A..90L}.

With our interior-atmosphere model in a MCMC framework, we obtained the posterior distribution of the compositional parameters (CMF and WMF) and the atmospheric parameters assuming a water-dominated volatile layer of each of the planets in these multiplanetary systems. We found that K2-138 and TRAPPIST-1 present a very clear volatile trend with distance from the host star. Kepler-102 could potentially present this trend. For the TOI-178 and Kepler-11 systems, our modelling ruled out the presence of large hydrosphere as responsible for their low density. For such systems, it would be necessary to include H/He as part of the volatile layer in a self-consistent interior-atmosphere model. Nonetheless, all multiplanetary systems showed that the volatile mass fraction is significantly lower for the inner planets than for the outer planets. This is consistent with a formation history that involves formation of the outer planets in the vicinity of the ice line, inwards migration and atmospheric loss of the inner planets. We discussed the possible formation and evolution pathways that might yield these volatile content trends case-by-case. Similarly, we also commented on the possible causes of the high core mass fractions of the inner planets of Kepler-102 and Kepler-80, which might involve formation in the vicinity of the rocklines. 

In addition, the atmospheric thickness that we obtained as a result of our Bayesian analysis (see Table~\ref{output_mcmc}) can be used to estimate the scale height of the extended atmospheres of the planets analysed in this work, which is necessary to assess the observing time and number of transits to characterise the composition of these atmospheres with transmission spectroscopy. This would confirm the exact composition of their atmospheres. To better assess possible evolutionary effects on the current composition of the planet, future work should involve the inclusion of atmospheric mass loss processes in the coupled atmosphere-interior model. In this work, we assumed that the planets do not evolve with time. The variation of water mass fraction could also have been shaped by post-formation processes such as hydrodynamic escape \citep{Bonfanti20}. Each of the discussed processes has been studied individually with interior models to constrain whether the atmospheres of low-mass planets are primordial or secondary \citep{dorn_heng18,gupta21}, but none has modelled the effects of all these combined processes on the volatile reservoir of low-mass planets.

\begin{acknowledgements}
We would like to thank Maria Bergemann and Matthew Raymond Gent for a preliminary analysis of the stellar spectrum. This research has made use of the services of the ESO Science Archive Facility.
This research was made possible through the use of the AAVSO Photometric All-Sky Survey (APASS), funded by the Robert Martin Ayers Sciences Fund. This publication makes use of data products from the Two Micron All Sky Survey, which is a joint project of the University of Massachusetts and the Infrared Processing and Analysis Center/California Institute of Technology, funded by the National Aeronautics and Space Administration and the National Science Foundation. This publication makes use of data products from the Wide-field Infrared Survey Explorer, which is a joint project of the University of California, Los Angeles, and the Jet Propulsion Laboratory/California Institute of Technology, funded by the National Aeronautics and Space Administration. 
This paper includes data collected by the K2 mission. Funding for the K2 mission is provided by the NASA Science Mission directorate.
This research has made use of the Exoplanet Follow-up Observation Program website, which is operated by the California Institute of Technology, under contract with the National Aeronautics and Space Administration under the Exoplanet Exploration Program.
This research has made use of NASA's Astrophysics Data System Bibliographic Services.
This work has made use of data from the European Space Agency (ESA) mission
{\it Gaia} (\url{https://www.cosmos.esa.int/gaia}), processed by the {\it Gaia}
Data Processing and Analysis Consortium (DPAC,
\url{https://www.cosmos.esa.int/web/gaia/dpac/consortium}). Funding for the DPAC
has been provided by national institutions, in particular the institutions
participating in the {\it Gaia} Multilateral Agreement.
This research has made use of the VizieR catalogue access tool, CDS, Strasbourg, France. The original description of the VizieR service was published in A\&AS, 143, 23.
TM acknowledges financial support from Belspo for contract PRODEX PLATO mission development. We acknowledge the anonymous referee whose comments helped improve and clarify this manuscript.
\end{acknowledgements}

\bibliographystyle{aa} 
\bibliography{bib} 

\begin{appendix}
\section{System parameters}
\newpage
\onecolumn
\begin{longtable}{lcc}
\caption{\label{MCMCprior} List of parameters used in the analysis. The priors are provided together with the posteriors. The posterior values represent the median and 68.3\% credible interval. Derived values that might be useful for follow-up work are also reported.} \\
\hline
Parameter & Prior & Posterior \\
\hline
\endfirsthead
\multicolumn{3}{l}{{\bfseries \tablename\ \thetable{} -- continued from previous page}} \\
\hline
Parameter & Prior & Posterior\\
\hline
\endhead
\multicolumn{3}{l}{{Continued on next page}} \\ 
\hline
\endfoot
\hline
\multicolumn{3}{l}{Notes:}\\
\multicolumn{3}{l}{$\bullet$ $\mathcal{N}(\mu,\sigma^{2})$: Normal distribution with mean $\mu$ and width $\sigma^{2}$}\\
\multicolumn{3}{l}{$\bullet$ $\mathcal{U}(a,b)$: Uniform distribution between $a$ and $b$}\\
\multicolumn{3}{l}{$\bullet$ $\mathcal{S}(a,b)$: Sine distribution between $a$ and $b$}\\
\multicolumn{3}{l}{$\bullet$ $\mathcal{T}(\mu,\sigma^{2},a,b)$: Truncated normal distribution with mean $\mu$ and width $\sigma^{2}$, between $a$ and $b$}\\
\endlastfoot
\\
\multicolumn{3}{l}{\it Stellar Parameters}\\
\\
Effective temperature \teff\ [K] & $\mathcal{N}(5275.0, 50.0)$ &  $5354.7^{_{+27.9}}_{^{-21.2}}$ \\
Surface gravity \logg\ [cgs] & $\mathcal{N}(4.5,0.11)$ & $4.55^{_{+0.02}}_{^{-0.02}}$  \\
Iron abundance \met\ [dex] &  $\mathcal{N}(0.08,0.05)$ & $0.07\pm0.05$ \\
Distance to Earth $D$ [pc] &  $\mathcal{N}(201.54,1.97)$ & $201.5\pm1.9$  \\
Interstellar extinction $E(B-V)$ [mag] &  $\mathcal{U}(0.0,1.0)$  & $0.006^{_{+0.009}}_{^{-0.005}}$ \\
Systemic radial velocity $\gamma$ [\kms] & $\mathcal{U}(-10.0,10.0)$  & $0.6392^{_{+0.0012}}_{^{-0.0013}}$ \\
Linear limb-darkening coefficient $u_{a}$ & (derived) & $0.4906^{_{+0.0075}}_{^{-0.0071}}$  \\
Quadratic limb-darkening coefficient $u_{b}$ & (derived) & $0.2084^{_{+0.0045}}_{^{-0.0047}}$  \\
Stellar density $\rho_{\star}/\rho_{\astrosun}$ & (derived) & $1.534^{_{+0.081}}_{^{-0.090}}$  \\
Stellar mass M$_{\star}$\ [\Msun] & (derived) & $0.891^{_{+0.017}}_{^{-0.027}}$  \\
Stellar radius R$_{\star}$\ [\Rsun] & (derived) & $0.834^{_{+0.011}}_{^{-0.01}}$ \\
Stellar age $\tau$\ [Gyr] & (derived) & $3.3^{_{+2.4}}_{^{-3.2}}$ \\
&& \\
\hline
\\
\multicolumn{3}{l}{\it Planet b Parameters}\\
\\
Orbital Period $P_{b}$ [d] &  $\mathcal{N}(2.35322,0.01)$ & $2.35308^{_{+0.00022}}_{^{-0.00023}}$  \\
Transit epoch $T_{0,b}$ [BJD - 2450000] &  $\mathcal{N}(7773.317,0.001)$ & $7773.31682^{_{+0.00092}}_{^{-0.00090}}$  \\
Radial velocity semi-amplitude $K_{b}$ [\kms] & $\mathcal{U}(0.0,0.1)$ & $0.00146^{_{+0.00049}}_{^{-0.00050}}$ \\
Orbital inclination $i_{b}$ [$^{\circ}$] & $\mathcal{S}(70.0,90.0)$ & $87.9^{_{+1.3}}_{^{-1.1}}$  \\
Planet-to-star radius ratio $k_{b}$ & $\mathcal{U}(0.0,1.0)$ & $0.01586^{_{+0.00072}}_{^{-0.00066}}$  \\
Orbital eccentricity $e_{b}$ & $\mathcal{T}(0.0,0.083,0.0,1.0)$ & $0.047^{_{+0.050}}_{^{-0.033}}$  \\
Argument of periastron $\omega_{b}$ [\degr] & $\mathcal{U}(0.0,360.0)$ & $169^{_{+93}}_{^{-109}}$ \\
System scale $a_{b}/R_{\star}$ & (derived) & $8.6^{_{+0.1}}_{^{-0.2}}$  \\
Impact parameter $b_{b}$ & (derived) & $0.305^{_{+0.175}}_{^{-0.191}}$  \\
Transit duration T$_{14,b}$ [h] & (derived) & $2.00^{_{+0.09}}_{^{-0.11}}$  \\
Semi-major axis $a_{b}$ [AU] & (derived) & $0.03332^{_{+0.00021}}_{^{-0.00034}}$  \\
Planet mass M$_{b}$ [\Mearth] & (derived) & $2.80^{_{+0.94}}_{^{-0.96}}$  \\
Planet radius R$_{b}$ [\Rearth] & (derived) & $1.442^{_{+0.071}}_{^{-0.063}}$  \\
Planet bulk density $\rho_{b}$ [\gcm3] & (derived) & $5.1^{_{+2.0}}_{^{-1.8}}$  \\
&& \\
\hline
\\
\multicolumn{3}{l}{\it Planet c Parameters}\\
\\
Orbital Period $P_{c}$ [d] &  $\mathcal{N}(3.55987,0.01)$ & $3.56004^{_{+0.00012}}_{^{-0.00011}}$  \\
Transit epoch $T_{0,c}$ [BJD - 2450000] &  $\mathcal{N}(7740.3223,0.001)$ & $7740.32185^{_{+0.00087}}_{^{-0.00090}}$ \\
Radial velocity semi-amplitude $K_{c}$ [\kms] & $\mathcal{U}(0.0,0.1)$ & $0.00270^{_{+0.00052}}_{^{-0.00051}}$  \\
Orbital inclination $i_{c}$ [$^{\circ}$] & $\mathcal{S}(70.0,90.0)$ & $88.7^{_{+0.8}}_{^{-0.7}}$  \\
Planet-to-star radius ratio $k_{c}$ & $\mathcal{U}(0.0,1.0)$ & $0.02418^{_{+0.00056}}_{^{-0.00051}}$  \\
Orbital eccentricity $e_{c}$ & $\mathcal{T}(0.0,0.083,0.0,1.0)$ & $0.037^{_{+0.041}}_{^{-0.025}}$ \\
Argument of periastron $\omega_{c}$ [\degr] & $\mathcal{U}(0.0,360.0)$ & $171^{_{+129}}_{^{-78}}$  \\
System scale $a_{c}/R_{\star}$ & (derived) & $11.3\pm0.2$  \\
Impact parameter $b_{c}$ & (derived) & $0.254^{_{+0.148}}_{^{-0.160}}$  \\
Transit duration T$_{14,c}$ [h] & (derived) & $2.37^{_{+0.05}}_{^{-0.06}}$  \\
Semi-major axis $a_{c}$ [AU] & (derived) & $0.04391^{_{+0.00028}}_{^{-0.00045}}$  \\
Planet mass M$_{c}$ [\Mearth] & (derived) & $5.95^{_{+1.17}}_{^{-1.12}}$  \\
Planet radius R$_{c}$ [\Rearth] & (derived) & $2.198^{_{+0.066}}_{^{-0.054}}$ \\
Planet bulk density $\rho_{c}$ [\gcm3] & (derived) & $3.1^{_{+0.7}}_{^{-0.6}}$  \\
&& \\
\hline
\\
\multicolumn{3}{l}{\it Planet d Parameters}\\
\\
Orbital Period $P_{d}$ [d] &  $\mathcal{N}(5.40478,0.01)$ & $5.40479\pm0.00021$  \\
Transit epoch $T_{0,d}$ [BJD - 2450000] &  $\mathcal{N}(7743.1607,0.001)$ & $7743.15984^{_{+0.00095}}_{^{-0.00093}}$ \\
Radial velocity semi-amplitude $K_{d}$ [\kms] & $\mathcal{U}(0.0,0.1)$ & $0.00285\pm0.00055$ \\
Orbital inclination $i_{d}$ [$^{\circ}$] & $\mathcal{S}(70.0,90.0)$ & $88.9^{_{+0.6}}_{^{-0.5}}$  \\
Planet-to-star radius ratio $k_{d}$ & $\mathcal{U}(0.0,1.0)$ & $0.02540^{_{+0.00069}}_{^{-0.00065}}$  \\
Orbital eccentricity $e_{d}$ & $\mathcal{T}(0.0,0.083,0.0,1.0)$ & $0.039^{_{+0.045}}_{^{-0.027}}$  \\
Argument of periastron $\omega_{d}$ [\degr] & $\mathcal{U}(0.0,360.0)$ & $207^{_{+69}}_{^{-138}}$ \\
System scale $a_{d}/R_{\star}$ & (derived) & $15.0\pm0.3$  \\
Impact parameter $b_{d}$ & (derived) & $0.297^{_{+0.145}}_{^{-0.170}}$  \\
Transit duration T$_{14,d}$ [h] & (derived) & $2.71^{_{+0.07}}_{^{-0.08}}$  \\
Semi-major axis $a_{d}$ [AU] & (derived) & $0.05800^{_{+0.00037}}_{^{-0.00059}}$  \\
Planet mass M$_{d}$ [\Mearth] & (derived) & $7.20^{_{+1.39}}_{^{-1.40}}$  \\
Planet radius R$_{d}$ [\Rearth] & (derived) & $2.310^{_{+0.077}}_{^{-0.068}}$  \\
Planet bulk density $\rho_{d}$ [\gcm3] & (derived) & $3.2\pm0.7$  \\
&& \\
\hline
\\
\multicolumn{3}{l}{\it Planet e Parameters}\\
\\
Orbital Period $P_{e}$ [d] &  $\mathcal{N}(8.26144,0.01)$ & $8.26146^{_{+0.00022}}_{^{-0.00021}}$  \\
Transit epoch $T_{0,e}$ [BJD - 2450000] &  $\mathcal{N}(7740.6451,0.001)$ & $7740.64563^{_{+0.00085}}_{^{-0.00087}}$  \\
Radial velocity semi-amplitude $K_{e}$ [\kms] & $\mathcal{U}(0.0,0.1)$ & $0.00387^{_{+0.00094}}_{^{-0.00093}}$  \\
Orbital inclination $i_{e}$ [$^{\circ}$] & $\mathcal{S}(70.0,90.0)$ & $88.7^{_{+0.3}}_{^{-0.2}}$  \\
Planet-to-star radius ratio $k_{e}$ & $\mathcal{U}(0.0,1.0)$ & $0.03604^{_{+0.00074}}_{^{-0.00072}}$ \\
Orbital eccentricity $e_{e}$ & $\mathcal{T}(0.0,0.083,0.0,1.0)$ & $0.049^{_{+0.048}}_{^{-0.034}}$  \\
Argument of periastron $\omega_{e}$ [\degr] & $\mathcal{U}(0.0,360.0)$ & $223^{_{+67}}_{^{-123}}$  \\
System scale $a_{e}/R_{\star}$ & (derived) & $19.8^{_{+0.3}}_{^{-0.4}}$  \\
Impact parameter $b_{e}$ & (derived) & $0.474^{_{+0.081}}_{^{-0.115}}$  \\
Transit duration T$_{14,e}$ [h] & (derived) & $2.97\pm0.05$  \\
Semi-major axis $a_{e}$ [AU] & (derived) & $0.07697^{_{+0.00050}}_{^{-0.00079}}$ \\
Planet mass M$_{e}$ [\Mearth] & (derived) & $11.28^{_{+2.78}}_{^{-2.72}}$ \\
Planet radius R$_{e}$ [\Rearth] & (derived) & $3.276^{_{+0.095}}_{^{-0.082}}$  \\
Planet bulk density $\rho_{e}$ [\gcm3] & (derived) & $1.8^{_{+0.5}}_{^{-0.4}}$  \\
&& \\
\hline
\\
\multicolumn{3}{l}{\it Planet f Parameters}\\
\\
Orbital Period $P_{f}$ [d] &  $\mathcal{N}(12.75759,0.01)$ & $12.75760^{_{+0.00051}}_{^{-0.00048}}$  \\
Transit epoch $T_{0,f}$ [BJD - 2450000] &  $\mathcal{N}(7738.7019,0.001)$ & $7738.70226^{_{+0.00093}}_{^{-0.00092}}$  \\
Radial velocity semi-amplitude $K_{f}$ [\kms] & $\mathcal{U}(0.0,0.1)$ & $0.00072^{_{+0.00091}}_{^{-0.00052}}$  \\
Orbital inclination $i_{f}$ [$^{\circ}$] & $\mathcal{S}(70.0,90.0)$ & $88.8^{_{+0.2}}_{^{-0.1}}$  \\
Planet-to-star radius ratio $k_{f}$ & $\mathcal{U}(0.0,1.0)$ & $0.03065^{_{+0.00085}}_{^{-0.00083}}$  \\
Orbital eccentricity $e_{f}$ & $\mathcal{T}(0.0,0.083,0.0,1.0)$ & $0.057^{_{+0.059}}_{^{-0.040}}$  \\
Argument of periastron $\omega_{f}$ [\degr] & $\mathcal{U}(0.0,360.0)$ & $172^{_{+117}}_{^{-112}}$  \\
System scale $a_{f}/R_{\star}$ & (derived) & $26.5\pm0.5$  \\
Impact parameter $b_{f}$ & (derived) & $0.541^{_{+0.073}}_{^{-0.109}}$  \\
Transit duration T$_{14,f}$ [h] & (derived) & $3.20\pm0.08$  \\
Semi-major axis $a_{f}$ [AU] & (derived) & $0.10283^{_{+0.00066}}_{^{-0.00105}}$ \\
Planet mass M$_{f}$ [\Mearth] & (derived) & $2.43^{_{+3.05}}_{^{-1.75}}$  \\
Planet radius R$_{f}$ [\Rearth] & (derived) & $2.787^{_{+0.093}}_{^{-0.085}}$  \\
Planet bulk density $\rho_{f}$ [\gcm3] & (derived) & $0.6^{_{+0.8}}_{^{-0.4}}$  \\
&& \\
\hline
\\
\multicolumn{3}{l}{\it Planet g Parameters}\\
\\
Orbital Period $P_{g}$ [d] &  $\mathcal{N}(41.97,0.1)$ & $41.96822^{_{+0.00817}}_{^{-0.00774}}$  \\
Transit epoch $T_{0,g}$ [BJD - 2450000] &  $\mathcal{N}(7773.76,2457773.93)$ & $7773.86006^{_{+0.01931}}_{^{-0.03522}}$  \\
Radial velocity semi-amplitude $K_{g}$ [\kms] & $\mathcal{U}(0.0,1.0)$ & $0.00049^{_{+0.00058}}_{^{-0.00035}}$  \\
Orbital inclination $i_{g}$ [$^{\circ}$] & $\mathcal{S}(70.0,90.0)$ & $89.5^{_{+0.4}}_{^{-0.3}}$  \\
Planet-to-star radius ratio $k_{g}$ & $\mathcal{U}(0.0,1.0)$ & $0.03199^{_{+0.00327}}_{^{-0.00248}}$  \\
Orbital eccentricity $e_{g}$ & $\mathcal{T}(0.0,0.083,0.0,1.0)$ & $0.054^{_{+0.060}}_{^{-0.038}}$  \\
Argument of periastron $\omega_{g}$ [\degr] & $\mathcal{U}(0.0,360.0)$ & $164^{_{+148}}_{^{-104}}$  \\
System scale $a_{g}/R_{\star}$ & (derived) & $58.6^{_{+1.0}}_{^{-1.2}}$  \\
Impact parameter $b_{g}$ & (derived) & $0.550^{_{+0.319}}_{^{-0.365}}$  \\
Transit duration T$_{14,g}$ [h] & (derived) & $4.71^{_{+0.79}}_{^{-1.63}}$  \\
Semi-major axis $a_{g}$ [AU] & (derived) & $0.22745^{_{+0.00146}}_{^{-0.00233}}$  \\
Planet mass M$_{g}$ [\Mearth] & (derived) & $2.45^{_{+2.92}}_{^{-1.74}}$  \\
Planet radius R$_{g}$ [\Rearth] & (derived) & $2.911^{_{+0.305}}_{^{-0.230}}$  \\
Planet bulk density $\rho_{g}$ [\gcm3] & (derived) & $0.5^{_{+0.7}}_{^{-0.4}}$  \\
&& \\
\hline
&& \\
\multicolumn{3}{l}{\it Instrument-related Parameters}\\
&& \\
HARPS jitter $\sigma_{j,~\rm RV}$ [\kms] & $\mathcal{U}(0.0,0.1)$ & $0.00146^{_{+0.00068}}_{^{-0.00077}}$  \\
\textit{K2} contamination [\%] & $\mathcal{T}(0.0,0.005,0.0,1.0)$ & $0.003^{_{+0.004}}_{^{-0.002}}$  \\
\textit{K2} jitter $\sigma_{j,~\textit{K2}}$ [ppm] & $\mathcal{U}(0.0, 10^5)$ & $185.9\pm2.7$  \\
\textit{K2} out-of-transit flux & $\mathcal{U}(0.99,1.01)$ & $1.0000058^{_{+0.0000037}}_{^{-0.0000038}}$ \\
SED jitter [mag]  & $\mathcal{U}(0.0,0.1)$ & $0.02^{_{+0.017}}_{^{-0.013}}$ \\
&& \\
\end{longtable}
\normalsize

\end{appendix}

\end{document}